# Ambient-Stable Transfer-Free Graphdiyne Wafers with Superhigh Hole Mobility at Room Temperature

**Beining Ma, Jianyuan Qi, Xinghai Shen***

*Corresponding Author: Prof. X. H. Shen
Fundamental Science on Radiochemistry and Radiation Chemistry Laboratory, Beijing National Laboratory for Molecular Sciences, Center for Applied Physics and Technology, College of Chemistry and Molecular Engineering, Peking University, Beijing 100871, P. R. China
E-mail: xshen@pku.edu.cn

Abstract: Graphdiyne (GDY) is recognized as a compelling candidate for the fabrication of next-generation high-speed low-energy electronic devices due to its inherent p-type semiconductor characteristics. However, the development of GDY for applications in field-effect transistors (FETs), complementary metal-oxide-semiconductor (CMOS), and logic devices remains constrained by the relatively low carrier mobility reported in current experimental studies. Herein, the synthesis of layer-controlled hydrogen-substituted graphdiyne (HsGDY) films directly on silicon substrates under a supercritical $CO_2$ atmosphere is reported, along with the fabrication of these films into HsGDY-based FETs. The transfer-free growth strategy eliminates performance degradation caused by post-synthesis transfer processes. The resulting HsGDY FETs exhibit a remarkable hole mobility of up to $3.8 \times 10^3$ $cm^2$ $V^{-1}$ $s^{-1}$ at room-temperature, which is an order of magnitude higher than that of most p-type semiconductors. The synthesis of transfer-free HsGDY wafers provides a new strategy for resolving the carrier mobility mismatch between p-channel and n-channel two-dimensional metal-oxide-semiconductor devices.

In the post-Moore era, silicon-based semiconductor technology faces critical challenges at sub-10 nm nodes, including short-channel effects, reduced carrier mobility, and increased power consumption.[1] As an emerging material, two-dimensional (2D) semiconductors are regarded as one of the most promising candidates for extending Moore's law due to their atomic thickness, ultra-flat surface, ultra-high carrier mobility and excellent electrical performance tunability.[2-4] Complementary transistors composed of p-type and n-type semiconductors are building blocks for 2D material-based integrated circuits.[5-6] Therefore, the performance of p-type semiconductors directly affects the behavior of complementary transistors. The full potential of two-dimensional materials remains largely constrained by the limited availability of p-type semiconductors with high carrier mobility and environmental stability.[7] At present, the carrier mobility of most p-type semiconductors typically ranges from approximately $10^{-2}$ to $10^2$ $cm^2$ $V^{-1}$ $s^{-1}$, and black phosphorus (BP) stands as the sole reported material exhibiting a room-temperature hole mobility exceeding $10^3$ $cm^2$ $V^{-1}$ $s^{-1}$.[8] The mismatch in carrier mobility between p-channel metal-oxide-semiconductor (PMOS) and n-channel metal-oxide-semiconductor (NMOS) device affects the speed of data processing significantly, therefore increasing power consumption and reducing performance.[9] Thus, it is of scientific and practical importance to develop p-type 2D semiconductors with wafer-scale synthesis and precise regulation of carrier mobility, so as to promote their applications in electronic devices such as complementary transistors, field-effect transistors (FETs) and logic gates.[10]

There are still many critical challenges to be overcome for exploring intrinsic p-type 2D semiconductors. Firstly, due to charge impurities and structural defects, most 2D semiconductors are either n-type or ambipolar.[11] Many potential electronic applications are limited by the lack of high-performance p-type semiconductors.[12] Secondly, realizing low-resistance ohmic contact to p-type semiconductors is crucial to avoid the reduction in switching speed and increase in power consumption caused by Schottky contact.[13] Finally, wafer-scale growth of p-type 2D semiconductors with uniform layer numbers and low-temperature (<400 °C) synthetic methods are needed.[7] In order to prepare high-density integrated devices with consistent performance, it is necessary to adjust and optimize the synthesis methods for different material systems to promote lateral growth while inhibiting vertical growth, so as to achieve uniform overall thickness.[7]

Graphdiyne (GDY), an emerging two-dimensional carbon allotrope,[14] features a planar structure comprising $sp^2$- and $sp$-hybridized carbon atoms.[15] GDY has attracted particular interest due to its physicochemical properties, such as large π-conjugated systems and well-distributed pore structures.[16] A variety of GDY-based novel devices have been developed in recent years, such as memories,[17-19] photodetectors[20,21] and artificial synapses.[22-24] Furthermore, GDY was theoretically predicted as an intrinsic p-type two-dimensional semiconductor with ultrahigh hole mobility.[25] It is expected to expand the types of p-type two-dimensional semiconductors and improve hole mobility. However, the experimental hole mobility of GDY field-effect transistors is relatively low,[26-28] which limits the application of GDY in CMOS. The transfer process becomes a critical bottleneck for enhancing the performance of GDY since it inevitably introduces impurity contamination and structural damage into the GDY films,[15] which degrades their hole mobility. In addition, the GDY film transferred onto silicon wafer is difficult to be photolithographically fabricated into FET arrays, which is a critical limitation for the application in electronic devices. To address these problems, it is imperative to develop a method for the *in situ* growth of two-dimensional GDY films on silicon wafer.

The production of transfer-free GDY wafer faces three challenges. Firstly, silicon substrates lack catalytic activity for the coupling of GDY monomers,[29] necessitating the introduction of an external catalytic source to drive the polymerization process. Secondly, the unrestricted rotation of aryl-alkyne groups and ethynylene linkages during the coupling process promotes out-of-plane growth, resulting in non-long-range order 3D architectures rather than the desired 2D crystalline lattices.[17,30,31] Finally, the strong interlayer van der Waals interactions and π-π stacking forces make GDY monolayers tend to stack into multilayers.[26,32] To date, researchers have successfully synthesized 2D GDY films on diverse substrates, including copper foils,[17] quartz,[29] and MXene matrices.[27] However, the *in situ* growth of 2D GDY films on silicon wafer surfaces remains unrealized.

Herein, we develop a synthetic methodology for the *in situ* growth of two-dimensional GDY films on silicon wafer surfaces via space-confined synthesis strategy in supercritical $CO_2$. By sandwiching a copper foil with a silicon wafer, the copper ions released from the copper foil can migrate and be adsorbed onto the silicon surface, thereby catalyzing the coupling of GDY monomers and enabling direct epitaxial growth on the substrate. By fabricating transfer-free HsGDY wafer, we aim to enhance its hole mobility and explore its potential for electronic applications in FETs, ultimately addressing critical challenges in p-type semiconductors.

## Preparation of transfer-free HsGDY wafers

HsGDY films can be synthesized within the gap formed by laminating silicon wafers and copper foils (Fig. 1a and Supplementary Fig. S2a). The synthesis mechanism of GDY in a supercritical $CO_2$ atmosphere has been studied in detail and will be published elsewhere. After a 24-hour reaction period, the final product is a 1×1 cm hydrogen-substituted graphdiyne (HsGDY) film grown on the silicon wafer (Fig. 1b and Supplementary Fig. S2b). The Raman peak at 970 cm$^{-1}$ corresponds to the characteristic signal of the silicon wafer,[33] and those observed at 1357 cm$^{-1}$, 1573 cm$^{-1}$, 1934 cm$^{-1}$, and 2212 cm$^{-1}$ are assigned to the D band, G band, -C≡C-Cu vibrational mode, and -C≡C-C≡C- stretching mode of HsGDY, respectively (Fig. 1c).[34] In conjunction with the four characteristic peaks of HsGDY, this confirms the *in situ* growth of a HsGDY film on the silicon wafer. In addition to the four characteristic peaks of HsGDY, a new peak emerges at 2721 cm$^{-1}$. This peak corresponds to the second-order overtone of the D band (1357 cm$^{-1}$),[35] indicating a higher degree of structural regularity in the synthesized HsGDY film.

X-ray photoelectron spectroscopy (XPS) was employed to characterize the HsGDY film grown on the silicon wafer (Fig. 1d). The oxygen signal originates from both the $SiO_2$ layer of the silicon wafer and the chemically adsorbed oxygen at the edges of HsGDY. Deconvolution of the asymmetric C 1s peak reveals four component peaks located at 284.8, 286.2, 287.7, and 288.7 eV, corresponding to carbon-carbon double bonds (C=C), carbon-carbon triple bonds (C≡C), carbon-oxygen single bonds (C-O), and carbon-oxygen double bonds (C=O), respectively (Fig 1e).[28]

Scanning electron microscopy (SEM) combined with Energy-dispersive X-ray spectroscopy (EDS) elemental analysis confirmed the high uniformity of the few-layer HsGDY film (Fig. 2a). The distribution of carbon elements revealed a clear boundary between the film and the bare silicon wafer (Fig. 2b). Additionally, the region covered by the few-layer GDY film partially attenuated the silicon signal, resulting in a distinct boundary in the Si elemental map that corroborates the C elemental distribution (Fig. 2c). These characterizations mutually validate the presence and uniformity of the HsGDY film on the silicon wafer. By precisely modulating the concentration of triethynylbenzene (TEB), HsGDY wafers with varying layer numbers can be successfully fabricated (Supplementary Figs. S3 and

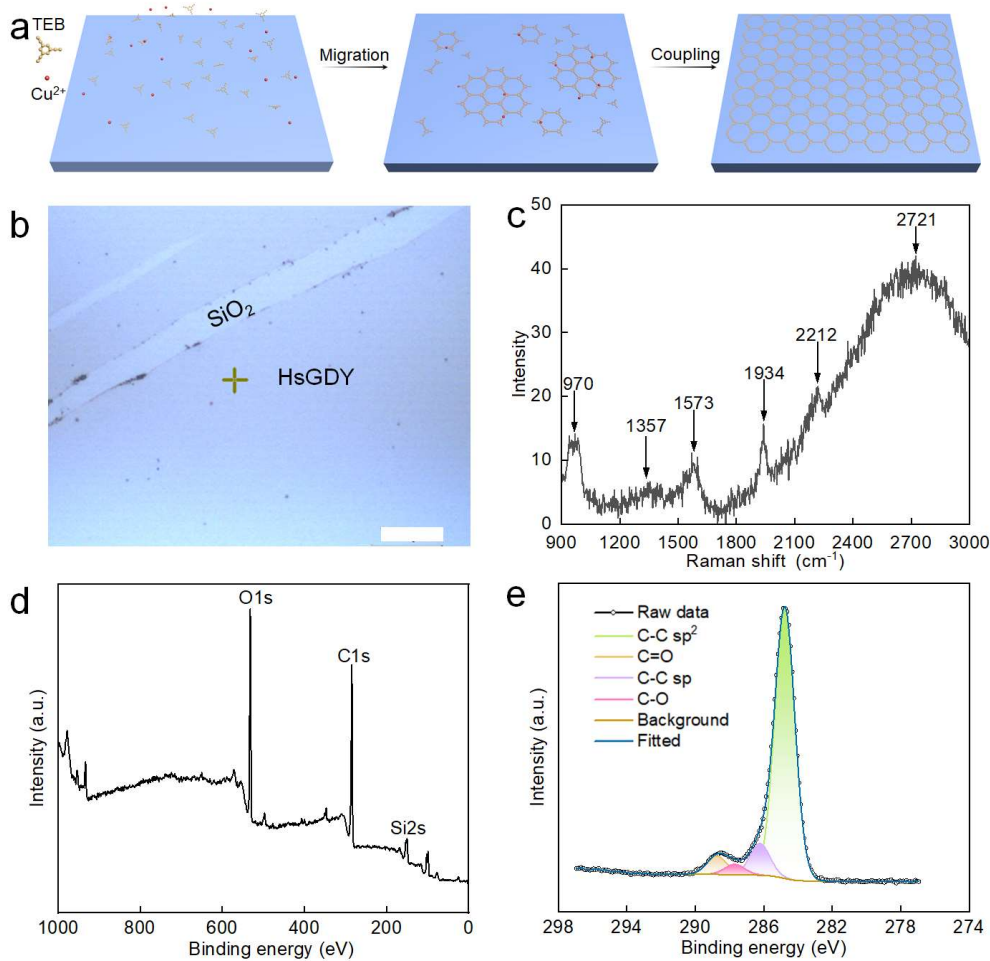

**Fig. 1 The *in situ* growth of HsGDY on SiO₂/Si. a**, Schematic of the growth mechanism of HsGDY on Si substrate. **b**, Optical image of HsGDY grown on SiO₂/Si. Scale bar is 100 μm. **c**, Raman spectra of HsGDY/SiO₂/Si. **d**, XPS spectra of HsGDY and **e**, Deconvoluted XPS spectra depicting the C1s band.

---

S4). Atomic force microscopy (AFM) characterization revealed that the thinnest HsGDY film grown on the silicon wafer has a thickness of approximately 2.2 nm, corresponding to a six-layer HsGDY wafer (Fig. 2d).[36] Transmission electron microscopy (TEM) characterization revealed distinct lattice fringes, demonstrating the high crystallinity of the synthesized HsGDY wafer (Fig. 2e). The observed lattice spacings of 0.22, 0.46, and 0.42 nm correspond to the (066) plane,[37] (110) plane,[38] and interlayer spacing of HsGDY,[39] respectively (Fig. 2f and Supplementary Fig. S5). These results confirm the well-ordered crystalline structure of the material.

## Photolithography processing of HsGDY FETs

In contrast to HsGDY FETs prepared by transfer process,[28] the *in situ* grown HsGDY films on silicon substrates can be integrated into standard photolithography workflows, enabling the fabrication of HsGDY FETs with precisely defined Hall bar geometries (Fig. 3a). The large-scale *in situ* grown HsGDY wafer demonstrates high processability for subsequent device fabrication, theoretically enabling the patterning of HsGDY films into arbitrary geometries via photolithography. To precisely evaluate the electrical properties of HsGDY, the as-prepared 6-layer wafer was fabricated into a standard Hall bar configuration, featuring a channel with dimensions of 110 μm in length and 20 μm in width (Fig. 3b). Following photolithographic alignment and magnetron sputtering processes, Au/Ti films (50 nm/5 nm) were deposited onto the edges of the Hall bar, serving as electrical contacts (Figs 3c and 3d). The device architecture integrates 8 electrodes, enabling comprehensive characterization of the electrical properties of HsGDY.

## Electronic properties of HsGDY FETs

Two-dimensional p-type semiconductors are considered critical for the development of high-speed

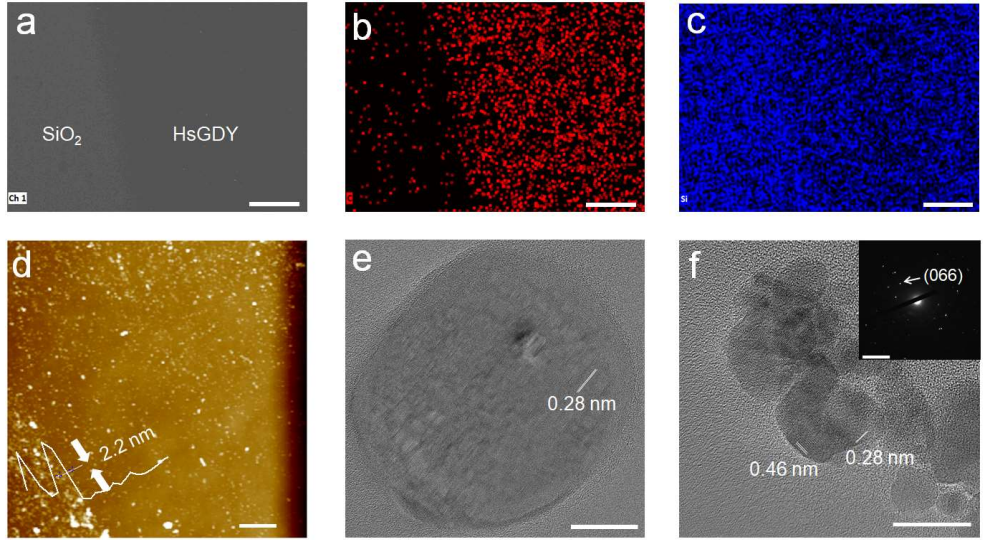

**Fig. 2 Characterization of the HsGDY wafer. a**, SEM image of the HsGDY wafer, showing a distinct boundary between the synthesized HsGDY and SiO$_2$ substrate. Scale bar is 50 μm. **b**, The element distribution map of carbon in **a**. Scale bar is 50 μm. **c**, The element distribution map of silicon in **a**. Scale bar is 50 μm. **d**, AFM image of the transfer-free HsGDY wafer. Scale bar is 1 μm. **e**, and **f**, TEM image of HsGDY and SAED pattern (inset). Scale bars of TEM images are both 20 nm, and scale bar of SAED pattern is 5 nm$^{-1}$.

---

and low-power electronic devices in the post-Moore era.[2] GDY, an intrinsic p-type semiconductor with a natural narrow bandgap, exhibits exceptional promise for such applications.[27] However, the currently reported hole mobilities of GDY-based devices remain substantially low,[26-28] which is primarily attributed to the structural damage and impurity contamination introduced during the transfer of GDY films onto silicon substrates for FET fabrication. The transfer-free HsGDY wafer developed in this work does not require a spin-coated PMMA layer for transfer to silicon substrates, thereby avoiding residual PMMA contamination and enhancing the carrier mobility of HsGDY.

The HsGDY FET was constructed by depositing Au/Ti on transfer-free HsGDY wafer as source and drain terminals, with a single-crystal Si substrate serving as the bottom gate and SiO$_2$ as the dielectric layer (Fig. 4a). The gate leakage current density of the fabricated HsGDY FET was below $10^{-3}$ A cm$^{-2}$ at ± 30 V, confirming the superior insulating properties of the SiO$_2$ dielectric layer (Supplementary Fig. S6). The $I_{ds}$-$V_{ds}$ curve measured at 298 K confirmed an Ohmic contact between HsGDY channel and the electrodes, with the conductivity of $2.3 \times 10^3$ S m$^{-1}$ (Fig. 4c).

The transfer characteristic curve of the 6-layer HsGDY FET, measured at $V_{ds}$ = -0.05 V, exhibits a sharp current surge at $V_g \approx$ -5 V (Fig. 4b). The observed current reduction during the forward sweep of $V_g$ indicates hole-dominated conduction, confirming the p-type nature of HsGDY (Fig. 4b).[26,27] Importantly, the HsGDY FET device maintains constant performance after being left in ambient air without any encapsulation for 60 days (Fig. 4b). Meanwhile, the output characteristic curves demonstrate an obvious gate control property, with distinct $I_{ds}$-$V_{ds}$ curves generated under varying gate biases (Fig. 4d). The carrier mobility of HsGDY was calculated using the equation[40] (1):

$$\mu = \left[\frac{dI_{ds}}{dV_g}\right]\left[\frac{L}{WC_gV_{ds}}\right]$$

where $L$ and $W$ represent the channel length and width, respectively, and the $C_g$ is 719.4 nF cm$^{-2}$. The calculated average hole mobility from three independent 6-layer HsGDY FET was $3.8 \times 10^3$ cm$^2$ V$^{-1}$ s$^{-1}$ (Supplementary Fig. S7 and Table S1). Four HsGDY FETs with different thicknesses were fabricated and characterized, demonstrating an increase in hole mobility from $7.3 \times 10^2$ cm$^2$ V$^{-1}$ s$^{-1}$ to $3.8 \times 10^3$ cm$^2$ V$^{-1}$ s$^{-1}$ as the film thickness decreased from 22 nm to 2.2 nm (Fig. 4e). Furthermore, the HsGDY FET shows a notable $I_{on}/I_{off}$ ratio of $1 \times 10^4$, indicating its potential applications of CMOS.

Critically, the room-temperature hole mobility of HsGDY is an order of magnitude higher than that of most p-type semiconductors.[41-49] The superhigh hole mobility observed in HsGDY FET is attributed to the elimination of impurity contamination and structural defects caused by transfer processes. To explain the mechanism underlying the p-type semiconducting behavior of HsGDY, electron paramagnetic resonance (EPR) measurements were conducted. HsGDY exhibits a symmetric EPR signal with a g-

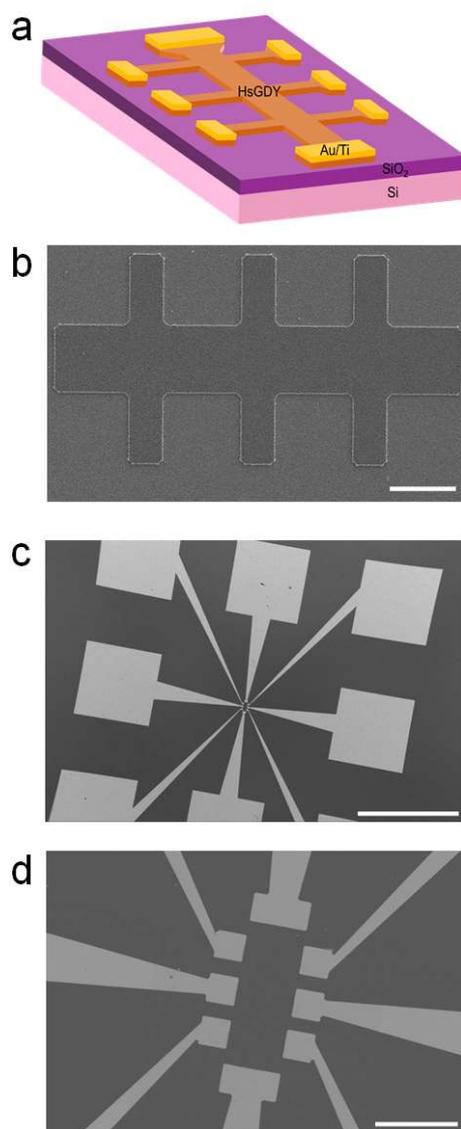

**Fig. 3 Photolithographic processing of HsGDY FET. a**, Schematic illustration of the HsGDY FET device structure. **b**, SEM image of the photolithographically defined HsGDY Hall bar structure, with dimensions of 110 μm×20 μm (length×width). Scale bar is 20 μm. **c**, SEM image of the HsGDY FET device structure. Scale bar is 1 mm. **d**, SEM image of the 8 electrical contacts. Scale bar is 50 μm.

---

value of 2.002 (Supplementary Fig. S10), characteristic of oxygen vacancies. In HsGDY, oxygen atoms are typically present as C=O groups at the edges. The removal of oxygen atoms disrupts the local conjugation, creating carbon dangling bonds. These dangling bonds function as strong electron acceptors, capturing electrons from the valence band and generating mobile holes in HsGDY. The directional movement of these holes under an applied electric field is responsible for the observed p-type conductivity in HsGDY.

The development of p-type 2D materials with high stability in air and carrier mobility is highly desirable and important. The HsGDY synthesized in this work functions as an intrinsic p-type semiconductor, exhibiting both high stability in air and high carrier mobility. This material holds significant implications for electronic devices, particularly CMOS technology, as its core impact is anticipated to substantially enhance device performance and energy efficiency. Specifically, HsGDY can be matched with high-carrier-mobility n-type semiconductors to fabricate heterogeneous integrated CMOS, thereby addressing the carrier mobility imbalance between PMOS and NMOS transistors. Furthermore, the increased hole mobility enables PMOS transistors to deliver a larger on-state current at identical drive voltages or to achieve faster switching speeds, leading to an increase in switching speed of fundamental CMOS logic gates. The increased processing speed of a single logic gate directly translates to higher maximum operating frequencies in integrated circuits. Finally, higher mobility permits transistors to achieve the required switching speed at lower operating voltages, which significantly reduces the energy consumption per logic operation.

## Conclusions

In this work, we demonstrated the *in situ* growth of few-layer HsGDY directly on silicon substrates under a $scCO_2$ atmosphere. By implementing a simple confinement approach, in which the Si wafer and Cu foil are simply sandwiched together, Si substrate can obtain large-area uniform few-layer HsGDY films. The transfer-free HsGDY wafers can be integrated into standard photolithography workflows, enabling the fabrication of HsGDY FETs with precisely defined Hall bar geometries The as-prepared HsGDY FETs exhibit both high hole mobility and ambient stability. Electrical characterization revealed p-type semiconducting behavior in HsGDY FETs with a hole mobility of $3.8\times10^3$ cm$^2$ V$^{-1}$ s$^{-1}$, surpassing all previously reported p-type 2D semiconductors. By preparing transfer-free HsGDY wafers, we have concurrently resolved two challenges in p-type 2D semiconductors: low hole mobility and wafer-scale thickness nonuniformity. Furthermore, HsGDY with high hole mobility is expected to be applied in CMOS to solve the problem of unbalanced mobility of n-type semiconductors and p-type semiconductors. By integrating an n-type semiconductor with matched carrier mobility and threshold voltage to HsGDY, the co-fabrication of CMOS devices is projected to achieve enhancement in data processing speeds and reduction in power consumption, thereby overcoming the performance limitations. The ability to adjust the hole mobility according to the thickness, combined with the fact that few-layer HsGDY has a

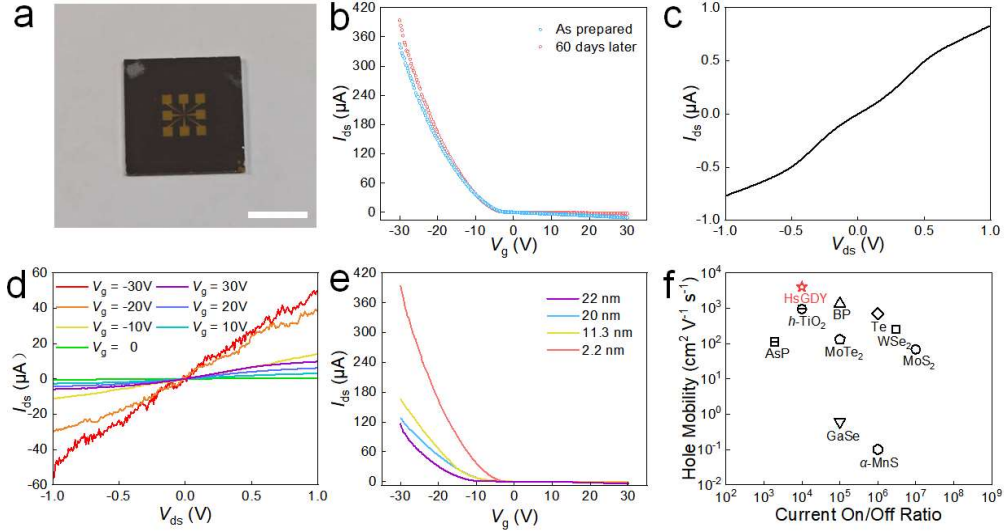

**Fig. 4 Electrical characterization of the HsGDY FETs. a**, Photograph of manufactured HsGDY FET. Scale bar is 5 mm. **b**, Transfer characteristic curve of the 6-layer HsGDY FET measured immediately and 60 days after fabrication. **c**, $I_{ds}$-$V_{ds}$ characteristic of the 6-layer HsGDY FET measured at 298 K. **d**, Output characteristic curves of the 6-layer HsGDY FET recorded under various $V_g$ biases from -30 to 30 V. **e**, Transfer characteristic curves of HsGDY FETs with different thicknesses. **f**, Comparisons of HsGDY FET with previously reported p-type 2D semiconductors in hole mobility.

---

direct bandgap, makes HsGDY a candidate for future nanoelectronic and optoelectronic applications.

## Methods

### Synthesis of HsGDY wafers

Si (100) wafers were used as growth substrates. These substrates were immersed in 100 °C piranha solution (35 mL $H_2SO_4$ + 15 mL $H_2O_2$), cleaned by ultrasonication in deionized water and dried with an $N_2$ gun to ensure thorough cleaning. The copper foil was electrochemical polished, followed by sequential washed with deionized water, ethanol and acetone. After cleaning, the copper foil was dried by $N_2$.

Si wafer was placed in the folded copper foil, ensuring that the smooth surface of the wafer was in direct contact with the copper foil. The assembled structure was placed flat at the bottom of a quartz basket, and a solution consisting of 70% TMEDA and 30% Pyr was added to the system. The quartz basket containing the assembled copper foil and Si wafer was pressurized with $scCO_2$ to reach a pressure of 100 bar. Subsequently, a solution of TEB in acetone (0.20 mg mL$^{-1}$) was pumped into the system. The reaction was carried out at 50 °C under light-protected conditions to prevent any undesired side reactions. After a reaction period of 24 hours, a 1×1 cm HsGDY film was successfully grown on the Si wafer. At the same monomer concentration, three HsGDY wafers were synthesized and fabricated as independent HsGDY FETs. The synthesized HsGDY wafers were immersed in 1 M HCl for 12 hours, and then rinsed with 10 % ammonia, 1 M HCl and deionized water in order to completely remove the residual copper species.

### Fabrication and measurements of HsGDY FETs

A 1×1 cm HsGDY wafer was spin-coated with photoresist, followed by sequential processing steps including baking, photolithography, development, rinsing, and reactive ion etching, yielding a Hall bar structure (110 μm in length×20 μm in width). Subsequent photolithographic alignment was performed, and Au/Ti electrodes (50 nm/5 nm) were deposited at the edges of the Hall bar via magnetron sputtering, thereby completing the fabrication of the HsGDY FET. The conductivity and transfer characteristics of the HsGDY FET were measured using a semiconductor device analyzer (Keysight Technologies B1500A).

The calculation of mobility is based on the following equations:

$$\mu = \left[\frac{dI_{ds}}{dV_g}\right]\left[\frac{L}{WC_gV_{ds}}\right] \quad (1)$$

$$C_g = \frac{\varepsilon_r\varepsilon_0}{d} \quad (2)$$

where $L$ and $W$ represent channel length and width, respectively. $C_g$ is the capacitance of the insulating layer, calculated by the dielectric constant of $SiO_2$ ($\varepsilon_r$), the vacuum dielectric constant ($\varepsilon_0$) and the thickness ($d$) of $SiO_2$ layer.

The calculation of conductivity is based on the following equation:

$$\sigma = \frac{dI}{dV}\frac{L}{WH} \quad (3)$$

where $L$, $W$ and $H$ represent channel length, width and thickness, respectively.

**Characterizations**

The obtained HsGDY wafers were subjected to Raman spectroscopy (Thermo-Fisher) using a 532 nm laser excitation source. The presence of characteristic Raman shifts corresponding to diyne bonds (-C≡C-C≡C-) was examined to confirm the existence of few-layer HsGDY on the Si wafer and to obtain structural information. X-ray photoelectron spectroscopy (XPS) studies were carried out using a ESCALAB250Xi (Thermo-Fisher) with monochromatic Al Kα radiation to investigate the chemical states of the samples. SEM analysis was performed using a Merlin Compact field-emission scanning electron microscope (ZEISS) to investigate the morphology and elemental distribution of the HsGDY film grown on the silicon wafer surface. Dimension FastscanBio atomic force microscope (Bruker) was used to perform the atomic force microscopy (AFM) measurements to analyze the thickness and uniformity of the samples. The transmission electron microscope (TEM) characterizations were performed using JEOL JEM-2100F with an operating voltage of 200 kV. The thickness of the $SiO_2$ dielectric layer was characterized using a UVISEL plus phase modulation ellipsometer (Horiba).

**Data availability**

The data that support the findings of this study are available in the supplementary material of this article.

**References**

1 Waldrop, M. M. The semiconductor industry will soon abandon its pursuit of Moore's law. Now things could get a lot more interesting. *Nature* **530**, 144-147 (2016).
2 Liu, Y. *et al.* Promises and prospects of two-dimensional transistors. *Nature* **591**, 43-53 (2021).
3 Fiori, G. *et al.* Electronics based on two-dimensional materials. *Nat. Nanotechnol.* **9**, 768-779 (2014).
4 Li, M. Y., Su, S. K., Wong, H. S. P. & Li, L. J. How 2D semiconductors could extend Moore's law. *Nature* **567**, 169-170 (2019).
5 Liu, Y., Huang, Y. & Duan, X. Van der Waals integration before and beyond two-dimensional materials. *Nature* **567**, 323-333 (2019).
6 Waltl, M. *et al.* Perspective of 2D integrated electronic circuits: scientific pipe dream or disruptive technology? *Adv. Mater.* **34**, 2201082 (2022).
7 Xiong, Y. *et al.* P-type 2D semiconductors for future electronics. *Adv. Mater.* **35**, e2206939 (2023).
8 Li, L. *et al.* Black phosphorus field-effect transistors. *Nat. Nanotechnol.* **9**, 372-377 (2014).
9 Kong, L., Chen, Y. & Liu, Y. Recent progresses of NMOS and CMOS logic functions based on two-dimensional semiconductors. *Nano Res.* **14**, 1768-1783 (2021).
10 He, Q. *et al.* Quest for p-type two-dimensional semiconductors. *ACS Nano* **13**, 12294-12300 (2019).
11 Zhao, T. *et al.* 2D $In_2Ge_2Te_6$ crystals for high-performance p-channel transistors. *Nano Lett.* **25**, 6235-6243 (2025).
12 Wang, Z., Nayak, P. K., Caraveo-Frescas, J. A. & Alshareef, H. N. Recent developments in p-type oxide semiconductor materials and devices. *Adv. Mater.* **28**, 3831-3892 (2016).
13 Wang, Y., Sarkar, S., Yan, H. & Chhowalla, M. Critical challenges in the development of electronics based on two-dimensional transition metal dichalcogenides. *Nat. Electron.* **7**, 638-645 (2024).
14 Li, G. *et al.* Architecture of graphdiyne nanoscale films. *Chem. Commun.* **46**, 3256-3258 (2010).
15 Gao, X., Liu, H., Wang, D. & Zhang, J. Graphdiyne: synthesis, properties, and applications. *Chem. Soc. Rev.* **48**, 908-936 (2019).
16 Fang, Y., Liu, Y., Qi, L., Xue, Y. & Li, Y. 2D graphdiyne: an emerging carbon material. *Chem. Soc. Rev.* **51**, 2681-2709 (2022).
17 Li, J. *et al.* Synthesis of wafer-scale ultrathin graphdiyne for flexible optoelectronic memory with over 256 storage levels. *Chem* **7**, 1284-1296 (2021).
18 Zhang, Z. *et al.* An ultrafast nonvolatile memory with low operation voltage for high-speed and low-power applications. *Adv. Funct. Mater.* **31**, 2102571 (2021).
19 Li, Y. *et al.* Low-voltage ultrafast nonvolatile memory via direct charge injection through a threshold resistive-switching layer. *Nat. Commun.* **13**, 4591 (2022).
20 Zhang, Y. *et al.* Graphdiyne-based flexible photodetectors with high responsivity and detectivity. *Adv. Mater.* **32**, 2001082 (2020).
21 Jin, Z. *et al.* Graphdiyne: ZnO nanocomposites for high-performance UV photodetectors. *Adv. Mater.* **28**, 3697-3702 (2016).
22 Hou, Y. *et al.* Large-scale and flexible optical synapses for neuromorphic computing and integrated visible information sensing memory processing. *Acs Nano* **15**, 1497-1508 (2021).
23 Yao, B. W. *et al.* Non-volatile electrolyte-gated transistors based on graphdiyne/$MoS_2$ with 23 robust stability for low-power neuromorphic computing and logic-in-memory. *Adv. Funct. Mater.* **31**, 2100069 (2021).
24 Wei, H. *et al*. Mimicking efferent nerves using a graphdiyne-based artificial synapse with multiple ion diffusion dynamics. *Nat. Commun.* **12**, 1-10 (2021).
25 Chen, J., Xi, J., Wang, D. & Shuai, Z. Carrier mobility in graphyne should be even larger than that in Graphene: A theoretical prediction. *J. Phys. Chem. Lett.* **4**, 1443-1448 (2013).
26 Zhou, J. *et al*. Synthesis of ultrathin graphdiyne film using a surface template. *ACS Appl. Mater. Interf.* **11**, 2632-2637 (2019).
27 Li, J. *et al*. Space-confined synthesis of monolayer graphdiyne in MXene interlayer. *Adv. Mater.* **36**, e2308429 (2024).
28 Josline, M. J. *et al*. Uniform synthesis of bilayer hydrogen substituted graphdiyne for flexible piezoresistive applications. *Small* **20**, e2307276 (2024).
29 Gao, X. *et al*. Direct synthesis of graphdiyne nanowalls on arbitrary substrates and its application for photoelectrochemical water splitting cell. *Adv. Mater.* **29**, 1605308 (2017).
30 Sakamoto, J., van Heijst, J., Lukin, O. & Schlueter, A. D. Two-dimensional polymers: Just a dream of synthetic chemists? *Angew. Chem., Int. Ed.* **48**, 1030-1069 (2009).
31 Zhou, J., Li, J., Liu, Z. & Zhang, J. Exploring approaches for the synthesis of few-layered graphdiyne. *Adv. Mater.* **31**, 1803758 (2019).
32 Colson, J. W. *et al*. Oriented 2D covalent organic framework thin films on single-layer graphene. *Science* **332**, 228-231 (2011).
33 Matsuoka, R. *et al*. Crystalline graphdiyne nanosheets produced at a gas/liquid or liquid/liquid interface. *J. Am. Chem. Soc.* **139**, 3145-3152 (2017).
34 Zhang, S. *et al*. Raman spectra and corresponding strain effects in graphyne and graphdiyne. *J. Phys. Chem. C* **120**, 10605-10613 (2016).
35 Beams, R., Cancado, L. G. & Novotny, L. Raman characterization of defects and dopants in graphene. *J. Phys-Condens. Mat.* **27** (2015).
36 Liu, R. *et al*. Chemical vapor deposition growth of linked carbon monolayers with acetylenic scaffoldings on silver foil. *Adv. Mater.* **29**, 1604665 (2017).
37 Lee, J., Li, Y., Tang, J. & Cui, X. Synthesis of hydrogen substituted graphyne through mechanochemistry and its electrocatalytic properties. *Acta Phys.-Chim. Sin.* **34**, 1080-1087 (2018).
38 Ren, X. *et al*. Tailoring acetylenic bonds in graphdiyne for advanced lithium storage. *ACS Sustain. Chem. Eng.* **8**, 2614-2621 (2020).
39 Lv, Q. *et al*. Selectively nitrogen-doped carbon materials as superior metal-free catalysts for oxygen reduction. *Nat. Commun.* **9** (2018).
40 Radisavljevic, B., Radenovic, A., Brivio, J., Giacometti, V. & Kis, A. Single-layer $MoS_2$ transistors. *Nat. Nanotechnol.* **6**, 147-150 (2011).
41 Chen, X. *et al*. High-quality sandwiched black phosphorus heterostructure and its quantum oscillations. *Nat. Commun.* **6**, 7315 (2015).
42 Wang, Y. *et al*. Field-effect transistors made from solution-grown two-dimensional tellurene. *Nat. Electron.* **1**, 228-236 (2018).
43 Liu, X. *et al*. P-type polar transition of chemically doped multilayer $MoS_2$ transistor. *Adv. Mater.* **28**, 2345-2351 (2016).
44 Li, N. *et al*. Synthesis and optoelectronic applications of a stable p-type 2D material: α-MnS. *ACS Nano* **13**, 12662-12670 (2019).
45 Fang, H. *et al*. High-performance single layered $WSe_2$ p-FETs with chemically doped contacts. *Nano Lett.* **12**, 3788-3792 (2012).
46 Late, D. J. *et al*. GaS and GaSe ultrathin layer transistors. *Adv. Mater.* **24**, 3549-3554 (2012).
47 Zhang, Q. *et al*. Simultaneous synthesis and integration of two-dimensional electronic components. *Nat. Electron.* **2**, 164-170 (2019).
48 Zhang, B. Y. *et al*. Hexagonal metal oxide monolayers derived from the metal-gas interface. *Nat. Mater.* **20**, 1073-1078 (2021).
49 Liu, B. *et al*. Black Arsenic-Phosphorus: layered anisotropic infrared semiconductors with highly tunable compositions and properties. *Adv. Mater.* **27**, 4423-4429 (2015).

## Acknowledgements

X.S. acknowledges the financial support from National Natural Science Foundation of China (grant no. U1830202). The authors thanked Molecular Materials and Nanofabrication Laboratory of Peking University for providing supports on device fabrication and characterizations.

## Author contributions

B.M. performed the experiments on the material synthesis, structure characterizations and device measurements. X.S. and B.M. cowrote the paper. All authors discussed the results and commented on the paper.

## Competing interests

The authors declare no competing interests.

## Supporting information

Supporting information is available from the author.

# Supporting Information

**Ambient-Stable Transfer-Free Graphdiyne Wafers with Superhigh Hole Mobility at Room Temperature**

*Beining Ma, Jianyuan Qi, Xinghai Shen*[*]

[*]Corresponding Author: Prof. X. H. Shen

Fundamental Science on Radiochemistry and Radiation Chemistry Laboratory, Beijing National Laboratory for Molecular Sciences, Center for Applied Physics and Technology, College of Chemistry and Molecular Engineering, Peking University, Beijing 100871, P. R. China

E-mail: xshen@pku.edu.cn

**Supplementary Information includes:**

**Supporting Methods**

**Supplementary Figures S1 to S10**

**Supplementary Tables S1 to S3**

**Supporting Methods**

**Materials**

All chemicals used in this work, including 1,3,5-triethynylbenzene (TEB), pyridine, tetramethylethylenediamine, petroleum ether (PE), dichloromethane (DCM), acetone concentrated hydrochloric acid, phosphoric acid, concentrated ammonia solution, absolute ethanol, ethylene glycol, N,N-dimethylformamide, high-purity nitrogen, carbon dioxide, and copper foil were purchased from suppliers and used without further purification unless otherwise noted. TEB was purified by column chromatography using PE and DCM as eluents. Deionized water (resistivity >18 MΩ·cm) was collected from a HHitech laboratory water purification system.

**TEM specimen preparation**

The specimens of TEM imaging experiments were prepared by ultrasonic exfoliation. The synthesized HsGDY wafer was immersed in 1-2 mL of absolute ethanol and subjected to ultrasonic treatment for 7 minutes. Four droplets of the supernatant were dropped onto a 300-mesh ultrathin copper grid and dried overnight under ambient conditions.

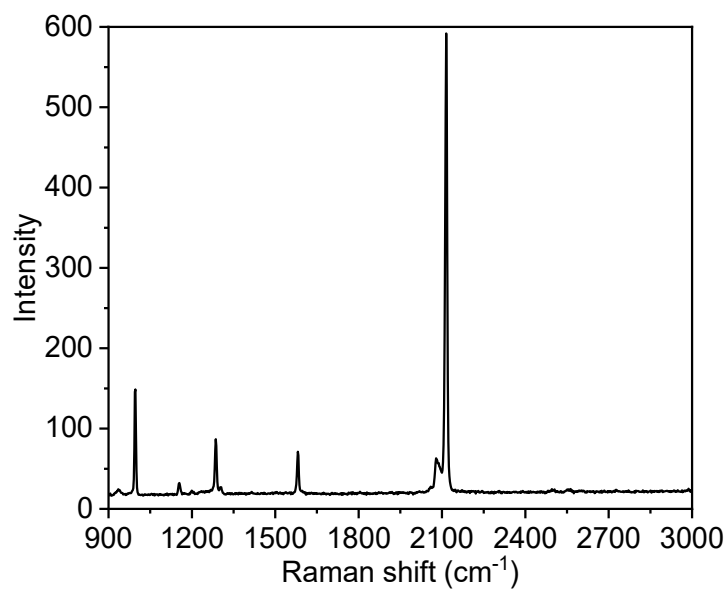

**Figure S1. The Raman spectra of purified TEB.** The peak observed at 996 cm$^{-1}$ is attributed to the silicon substrate. The Raman peaks at 1286 cm$^{-1}$, 1581 cm$^{-1}$, and 2115 cm$^{-1}$ are assigned to the D band, G band, and the ν(C≡C) stretching mode of the alkyne group, respectively.[1]

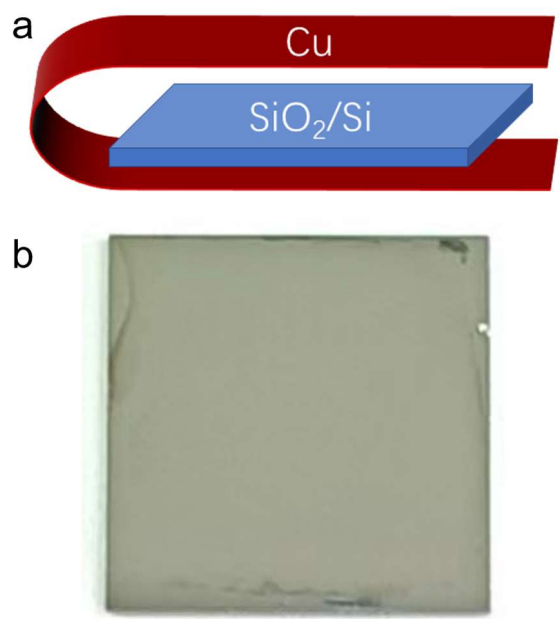

**Figure S2. a**, The schematic diagram of the relative position of copper foil and silicon wafer. **b**, The photo of transfer-free HsGDY wafer. The wafer size is 1 × 1 cm.

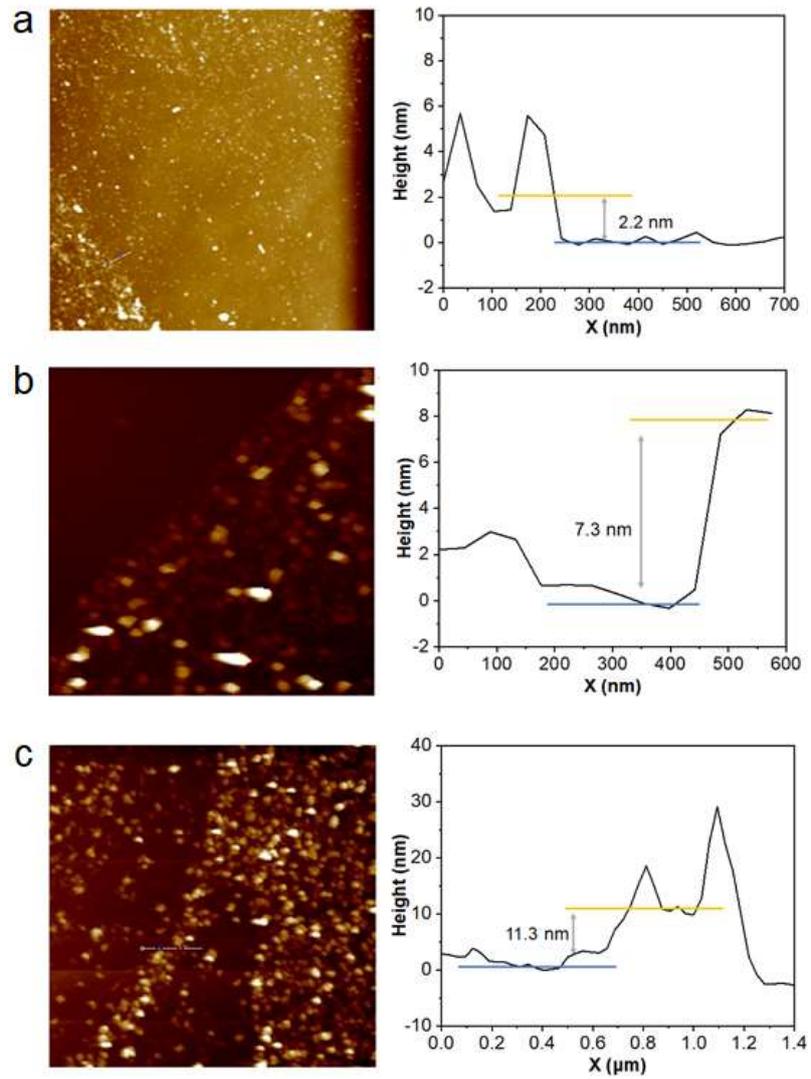

**Figure S3. The AFM images of transfer-free HsGDY wafers.** The monomer concentrations are (**a**) 0.20, (**b**) 0.22 and (**c**) 0.24 mg mL$^{-1}$, respectively.

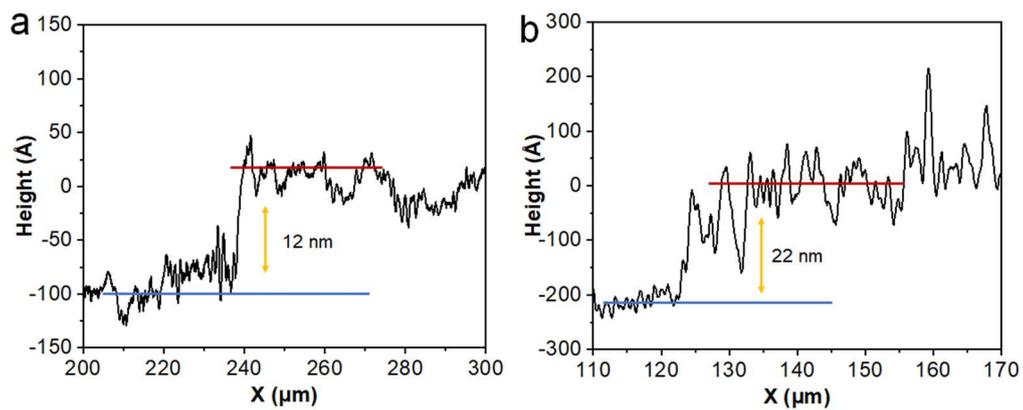

**Figure S4. Thickness profiles of transfer-free HsGDY wafers synthesized at monomer concentrations of 0.26 mg mL$^{-1}$ (a) and 0.32 mg mL$^{-1}$ (b).** The thickness profiles are measured using a stylus profilometer.

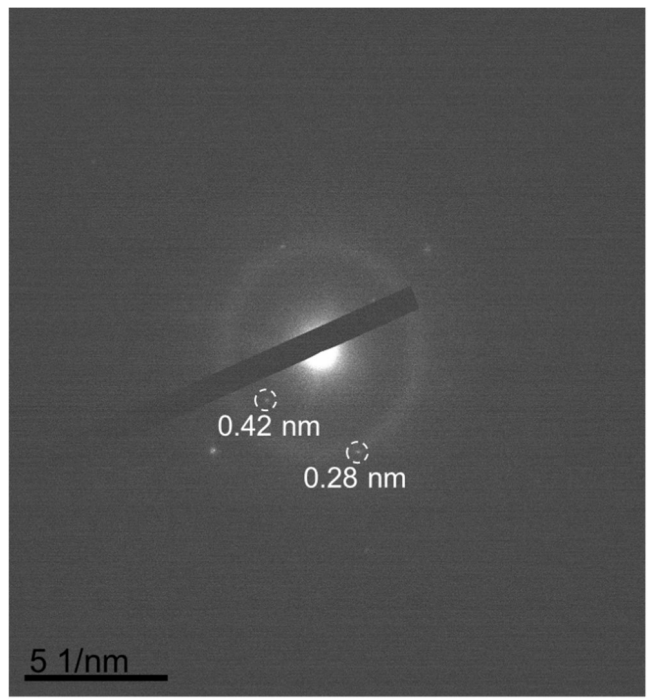
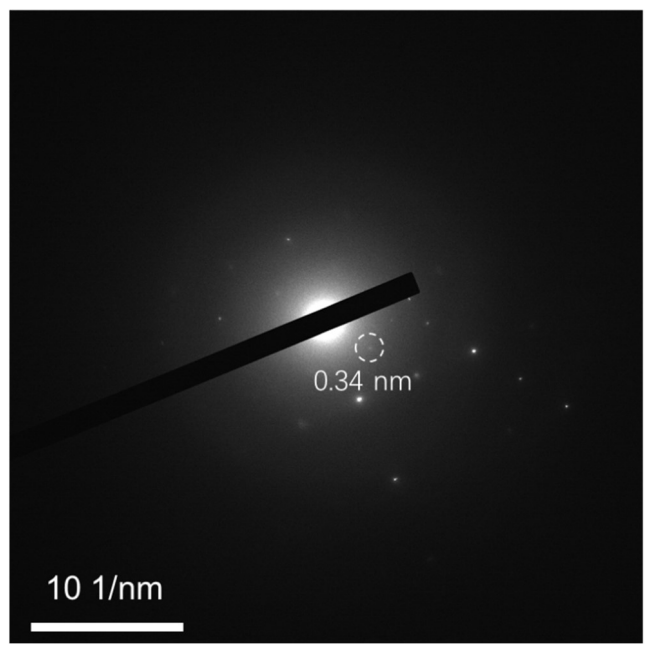

**Figure S5. Selected-area electron diffraction (SAED) pattern of the HsGDY films at two different areas.** The lattice spacings of 0.28 nm, 0.34 nm, and 0.42 nm are observed.

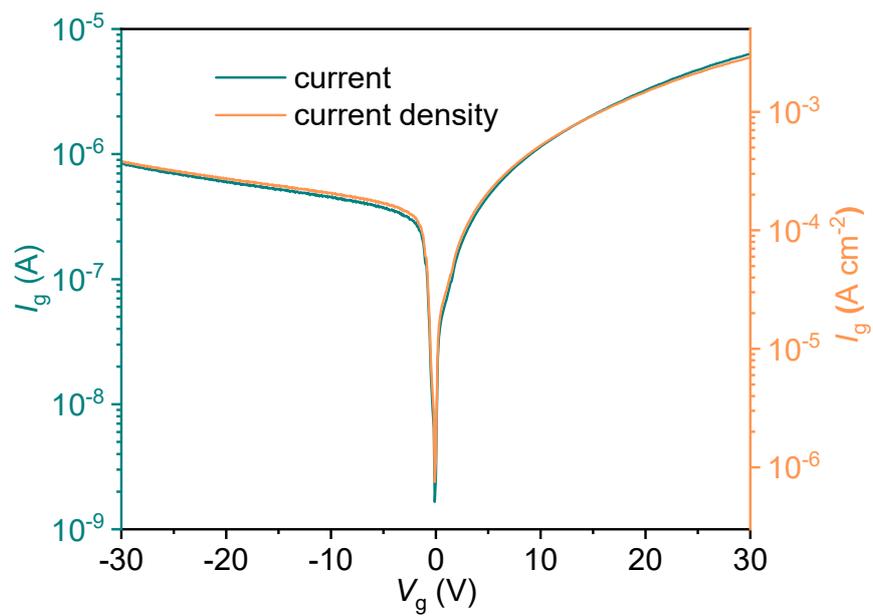

**Figure S6.** Leakage current of $SiO_2$ dielectric layer in the HsGDY FET.

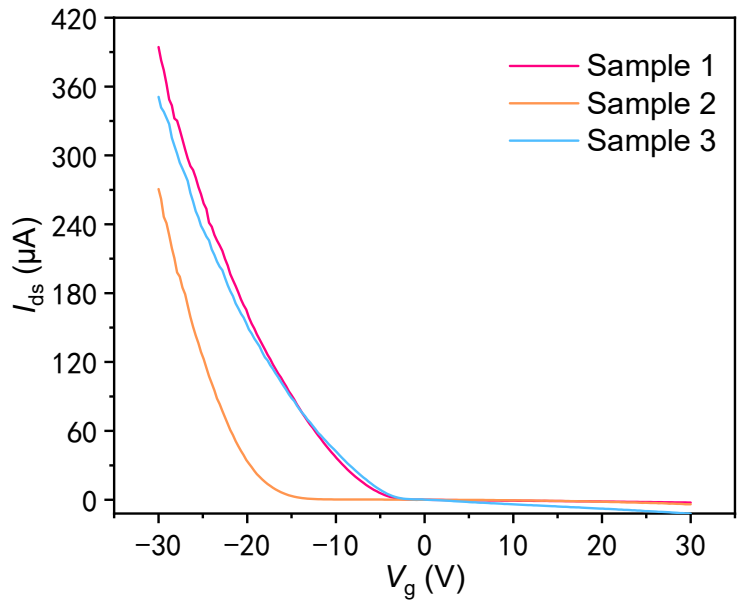

**Figure S7.** Transfer characteristic curves of the three independent 6-layer HsGDY FETs.

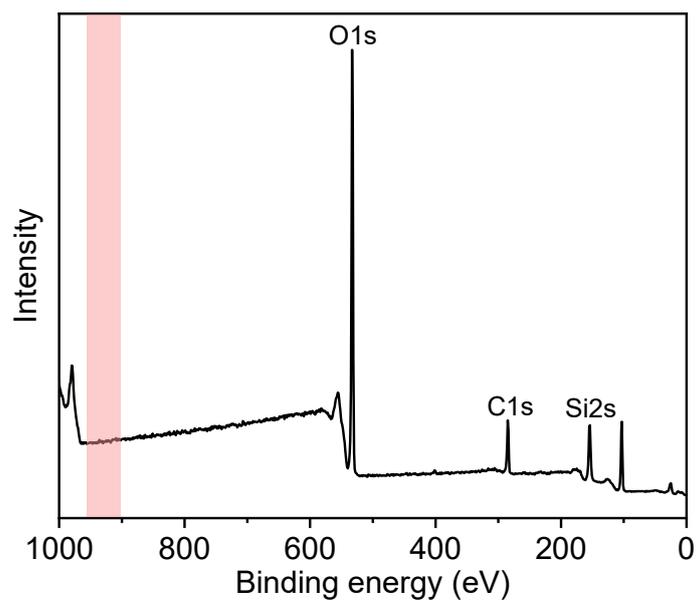

**Figure S8. XPS spectra of transfer-free HsGDY wafer washed by 1 M HCl.** The Cu signal peak in the red region disappears, indicating that the residual copper species are removed.

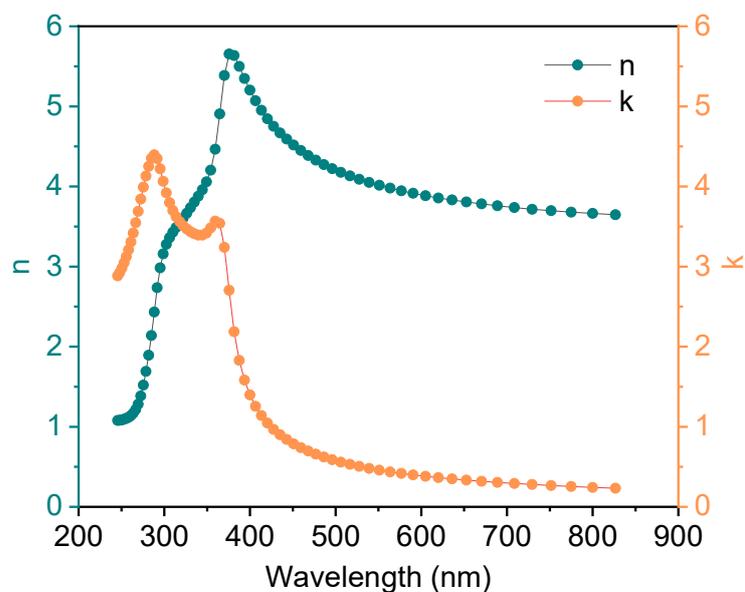

**Figure S9. Spectroscopic ellipsometry characterization of silicon wafers for HsGDY growth.** Measurements are performed over a wavelength range of from 245.5 to 826.7 nm, yielding a fitted $SiO_2$ layer thickness of 4.8 nm. At a specific wavelength of 496 nm, the complex refractive index components n and k are determined to be 4.22 and 0.59, respectively. The fitting procedure achieving a root-mean-square error of 0.035.

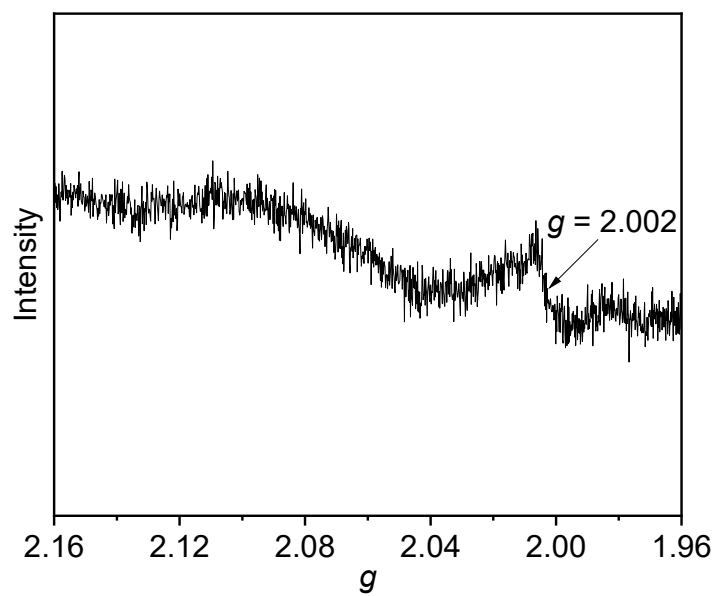

**Figure S10. Electron paramagnetic resonance spectra of HsGDY films grown in situ on silicon wafers.** HsGDY exhibits a symmetric EPR signal with a g-value of 2.002, which is a characteristic signal of oxygen vacancies.

**Table S1. Mobilities of three independent 6-layer HsGDY FETs.**

| Number of samples | d$I_{ds}$/d$V_g$ ($10^{-5}$ A V$^{-1}$) | Mobility ($10^3$ cm$^2$ V$^{-1}$ s$^{-1}$) |
|---|---|---|
| 1 | 2.56 | 3.9 |
| 2 | 2.63 | 4.0 |
| 3 | 2.30 | 3.5 |

**Table S2.** Comparison of semiconducting properties of HsGDY and p-type 2D materials in literature.

| p-type 2D Materials | Mobility ($cm^2$ $V^{-1}$ $s^{-1}$) | Ref. |
|---|---|---|
| HsGDY (in situ grown on $SiO_2$/Si) | 3800 | This work |
| Black Phosphorus | 1350 | 2 |
| Te | 700 | 3 |
| Ge | 724 | 4 |
| $MoS_2$ | 68 | 5 |
| $\alpha$-MnS | 0.1 | 6 |
| $WSe_2$ | 250 | 7 |
| GaSe | 0.6 | 8 |
| $MoTe_2$ | 130 | 9 |
| $h$-$TiO_2$ | 950 | 10 |
| CuO | 0.8 | 11 |
| AsP | 110 | 12 |

**Table S3. Comparison of mobility of GDY FET with similar reported devices.**

| GDY Materials | Thickness (nm) | Mobility ($cm^2 V^{-1} s^{-1}$) | Ref. |
| --- | --- | --- | --- |
| HsGDY Wafer | 2.2 | 3800 | This work |
| HsGDY(grown on Ge substrate) | 1.4 | 52.6 | 13 |
| Monolayer GDY | 0.78 | 247.1 | 14 |
| GDY/MoS$_2$ | 8.23 | 9.044 | 15 |
| GDY/GQD | 6 | 0.033 | 16 |
| GDY (homogenous reaction) | 67 | 13.3 | 17 |
| GDY (microwave) | 1 | 50.1 | 18 |
| GDY (surface-templated) | 3 | 6.3 | 19 |
| GDY/PFC | 7000 | 0.69 | 20 |
| GDY (VLS method) | 22 | 100 | 21 |

**Supporting References**

1 Zhang, S. *et al.* Raman spectra and corresponding strain effects in graphyne and graphdiyne. *J. Phys. Chem. C* **120**, 10605-10613 (2016).
2 Chen, X. *et al.* High-quality sandwiched black phosphorus heterostructure and its quantum oscillations. *Nat. Commun.* **6**, 7315 (2015).
3 Wang, Y. *et al.* Field-effect transistors made from solution-grown two-dimensional tellurene. *Nat. Electron.* **1**, 228-236 (2018).
4 Lan, H. H. *et al.* Self-limiting synthesis of ultrathin Ge(110) single crystal via liquid metal. *Small* **18**, 2106341 (2022).
5 Liu, X. *et al.* P-type polar transition of chemically doped multilayer $MoS_2$ transistor. *Adv. Mater.* **28**, 2345-2351 (2016).
6 Li, N. *et al.* Synthesis and optoelectronic applications of a stable p-type 2D material: α-MnS. *ACS Nano* **13**, 12662-12670 (2019).
7 Fang, H. *et al.* High-performance single layered $WSe_2$ p-FETs with chemically doped contacts. *Nano Lett.* **12**, 3788-3792 (2012).
8 Late, D. J. *et al.* GaS and GaSe ultrathin layer transistors. *Adv. Mater.* **24**, 3549-3554 (2012).
9 Zhang, Q. *et al.* Simultaneous synthesis and integration of two-dimensional electronic components. *Nat. Electron.* **2**, 164-170 (2019).
10 Zhang, B. Y. *et al.* Hexagonal metal oxide monolayers derived from the metal-gas interface. *Nat. Mater.* **20**, 1073-1078 (2021).
11 Liu, A. *et al.* Water-induced scandium oxide dielectric for low-operating voltage n- and p-type metal-oxide thin-film transistors. *Adv. Funct. Mater.* **25**, 7180-7188 (2015).
12 Liu, B. *et al.* Black Arsenic-Phosphorus: layered anisotropic infrared semiconductors with highly tunable compositions and properties. *Adv. Mater.* **27**, 4423-4429 (2015).
13 Josline, M. J. *et al.* Uniform synthesis of bilayer hydrogen substituted graphdiyne for flexible piezoresistive applications. *Small* **20**, e2307276 (2024).
14 Li, J. *et al.* Space-confined synthesis of monolayer graphdiyne in MXene interlayer. *Adv. Mater.* **36**, e2308429 (2024).
15 Do, D. P. *et al.* Highly efficient van der waals heterojunction on graphdiyne toward the high-performance photodetector. *Adv. Sci.* **10**, e2300925 (2023).
16 Ghafary, Z., Salimi, A., Hallaj, R., Akhtari, K. & Ghasemi, F. Light triggering performance of the van der Waals heterojunction of 2D/0D graphdiyne/graphdiyne quantum dot as a novel phototransistor. *Carbon* **215**, 118475 (2023).
17 Li, Y. *et al.* Light and heat triggering modulation of the electronic performance of a graphdiyne-based thin film transistor. *J. Phys. Chem. Lett.* **11**, 1998-2005 (2020).
18 Yin, C. *et al.* Catalyst-free synthesis of few-layer graphdiyne using a microwave-induced temperature gradient at a solid/liquid interface. *Adv. Funct. Mater.* **30**, 2001396 (2020).
19 Zhou, J. *et al.* Synthesis of ultrathin graphdiyne film using a surface template. *ACS Appl. Mater. Interf.* **11**, 2632-2637 (2019).
20 Cui, W. *et al.* High-performance field-effect transistor based on novel conjugated p-*o*-fluoro-p-alkoxyphenyl-substituted polymers by graphdiyne doping. *J. Phys. Chem. C* **121**, 23300-23306 (2017).
21 Qian, X. *et al.* Self-catalyzed growth of large-area nanofilms of two-dimensional carbon. *Sci. Rep.* **5**, 7756 (2015).